# Multi-hour stable trapping and threshold steady state of non-laser-coolable ions with ultracold atoms in a hybrid trap

Satyabrata Baidya[1] and Sourav Dutta[1,*]

[1] *Tata Institute of Fundamental Research, 1 Homi Bhabha Road, Colaba, Mumbai 400005, India*

We report multi-hour stable trapping and steady state of non-laser-coolable $Cs^+$ ions in a linear Paul trap, achieved via resonant charge-exchange (RCE) cooling with a precisely centered ultracold Cs cloud. Without cold atoms, all ions are lost within 3 minutes. With centered cold atoms, RCE cooling establishes a stable population of ions for 6 hours with no measurable decay – more than two orders of magnitude longer hold times compared to prior hybrid atom-ion systems. We observe a threshold behavior in the steady-state number of ions ($N_s$): initial ion loadings above $N_s$ ions converge downward to $N_s$ ions and remain constant thereafter; initial loadings below $N_s$ ions show no measurable decay. The dynamics is governed by a competition between ion-ion rf heating and ion-atom collisional cooling. The prolonged, simultaneous ion-atom trapping suggests an upper bound on the three-body recombination-induced $\mathrm{Cs}_2^+$ formation rate constant: $k_3 \ll 1.4 \times 10^{-25}\mathrm{cm}^6\mathrm{s}^{-1}$. This remarkable realization of long-lived steady state overcomes a critical limitation of prior hybrid atom-ion systems and enables extended studies of ultracold ion-neutral chemistry, rare inelastic collisions, and sympathetic cooling routes for complex molecular ions.

Stable, long-term trapping of ions has been the foundation for precision measurements, atomic clocks, and quantum information processing. Over the past two decades, hybrid atom-ion traps have emerged as a powerful platform to extend these capabilities [1], enabling studies of ultracold ion-atom collisions [2–7], sympathetic cooling [8–14], nonequilibrium dynamics [15–17], complex formation [18,19], spin-exchange [20] and charge-exchange reactions [2,5,21]. Despite remarkable progress, prolonged simultaneous trapping of ion ensembles and ultracold atoms has remained limited to transient regimes i.e. seconds to a few minutes, primarily due to rapid ion loss from the Paul-trap via charge exchange or micromotion heating, and/or atom loss from optical dipole traps (ODTs) via collisional heating. The short lifetime limits the accumulation of collision statistics, in-situ molecular formation, and systematic exploration of rare processes. Another limitation common to previous studies of atom-mediated ion cooling is the deliberate use of a single trapped ion to eliminate ion-ion collisional heating or to operate in a regime where Coulomb interactions are negligible. Although this approach facilitates experimental control and simplifies data interpretation, it precludes the benefits of ensemble averaging and prevents exploration of the rich dynamics arising from the interplay between ion-atom and ion-ion interactions in a dynamic trap. Here, we break through these longstanding barriers by realizing the long-lived steady state regime for ion ensembles with ultracold atoms, enabled by resonant charge exchange (RCE) cooling [11,22] which provides an exceptionally efficient velocity-reset mechanism where a single collision can transfer nearly all the ion's excess kinetic energy to the neutral atom, resulting is a cold ion.

We demonstrate 6 hours of stable trapping for ensembles of trapped $Cs^+$ ions in a linear Paul trap (LPT) with ultracold Cs atoms in a magneto-optical trap (MOT), thus realizing the long-lived steady state regime for ion ensembles that cannot be laser cooled. We report a genuine threshold behavior in the number of trapped ions at steady state ($N_s$) that arises primarily from competition between rf-induced heating and ion-atom sympathetic cooling. The steady state population $N_s$ and the lifetime increase with the number of ultracold atoms and decrease when the atomic cloud is displaced from the center of the LPT. The former is a result of increased ion-atom collision rate leading to faster sympathetic cooling while the latter is a result of increased micromotion induced heating. The results indicate that RCE cooling can actively balance rf micromotion heating and ion-ion rf heating, creating a steady state for an ensemble of non-laser-coolable ions with ultracold atoms. The orders-of-magnitude longer hold-time for ion ensembles demonstrated here enables both ensemble and time averaging for dramatically improved signal-to-noise in collision studies.

Our experimental apparatus is described in detail in Ref. [23] and the relevant parts are discussed in the supplementary file [24]. Briefly, we prepare $\sim 10^8$ ultracold Cs atom with a typical $1/e$ radius of ~0.9 mm at a temperature of ~150 μK in a three-dimensional (3D) magneto-optical trap (MOT) loaded from a differentially pumped 2D MOT. The loading of the 3D MOT is controlled using optical shutters. The $Cs^+$ ions are produced by photo-ionizing a small fraction of the Cs atoms in the $6p_{3/2}$ state using a short pulse of 505-nm light, the duration of which determines the number of $Cs^+$ ions. The $Cs^+$ ions are trapped in a linear Paul trap consisting of four rod electrodes (center-to-surface distance 13 mm) applied with radiofrequency (rf) voltages and two annular endcap electrodes (distance 88 mm) applied with dc voltages. Unless stated otherwise, the dc endcap voltages are kept fixed at +40V and the rf voltage ($V_{\mathrm{rf}}$) and frequency ($\Omega_{\mathrm{rf}}$) are 130V and 470 kHz, respectively. The radial and axial secular frequencies are measured to be 37 kHz and 12 kHz, respectively, using the parametric resonant quenching method, and are found to be in excellent agreement with ion trajectory simulations. The trap depth, $U \approx 0.9$ eV, is estimated from simulations [24,25]. To detect the $Cs^+$ ions, one of the endcap voltages is switched from +40 V to -10 V to extract the ions axially and project them onto a channel electron multiplier (CEM) where individual ions are detected as current spikes which are counted to determine the number of ions. The vacuum chamber is maintained at a pressure of $10^{-11}$ mbar.

*sourav.dutta@tifr.res.in

In the experiment, we load the LPT with $Cs^+$ ions and hold them for predetermined hold-times either in presence or absence of the ultracold Cs atoms. After the hold-time has elapsed, we extract the ions and count the number of ions detected in the CEM. We plot the measured number of ions $N(t)$ vs hold-time $t$ as shown in Fig. 1. In absence of the ultracold atoms, the number of ions in the LTP decreases and the LTP empties out within 180 s (squares). In contrast, in the presence of ultracold atoms, after an initial phase of ion loss, the number of ions in the LPT stabilizes and a sizable fraction of the ions remain trapped for 6 hours (circles). We interpret the data as evidence of ion cooling and realization of an ion-atom steady state as discussed in the following paragraph. To provide further evidence that ions are cooled by the ultracold atoms, we perform additional experiments where, after holding the ions in the LPT for a predetermined hold-time of 60 s, we suddenly reduce the trap depth of the LTP, hold the ions for a brief period of 100 ms, and then extract and detect the ions. The rationale is that the relatively hotter ions will escape from the LPT during the 100 ms of reduced trap depth while colder ions will remain trapped and be detected. In the inset of Fig. 1 we plot the normalized number of ions detected as function of the reduction ($\Delta U$) in the trap depth. When the trap depth is reduced by >0.2 eV, a higher number of ions is detected in the presence of ultracold atoms (circles) as opposed when atoms are absent (squares), thus providing a clear signature of ion cooling by the atoms.

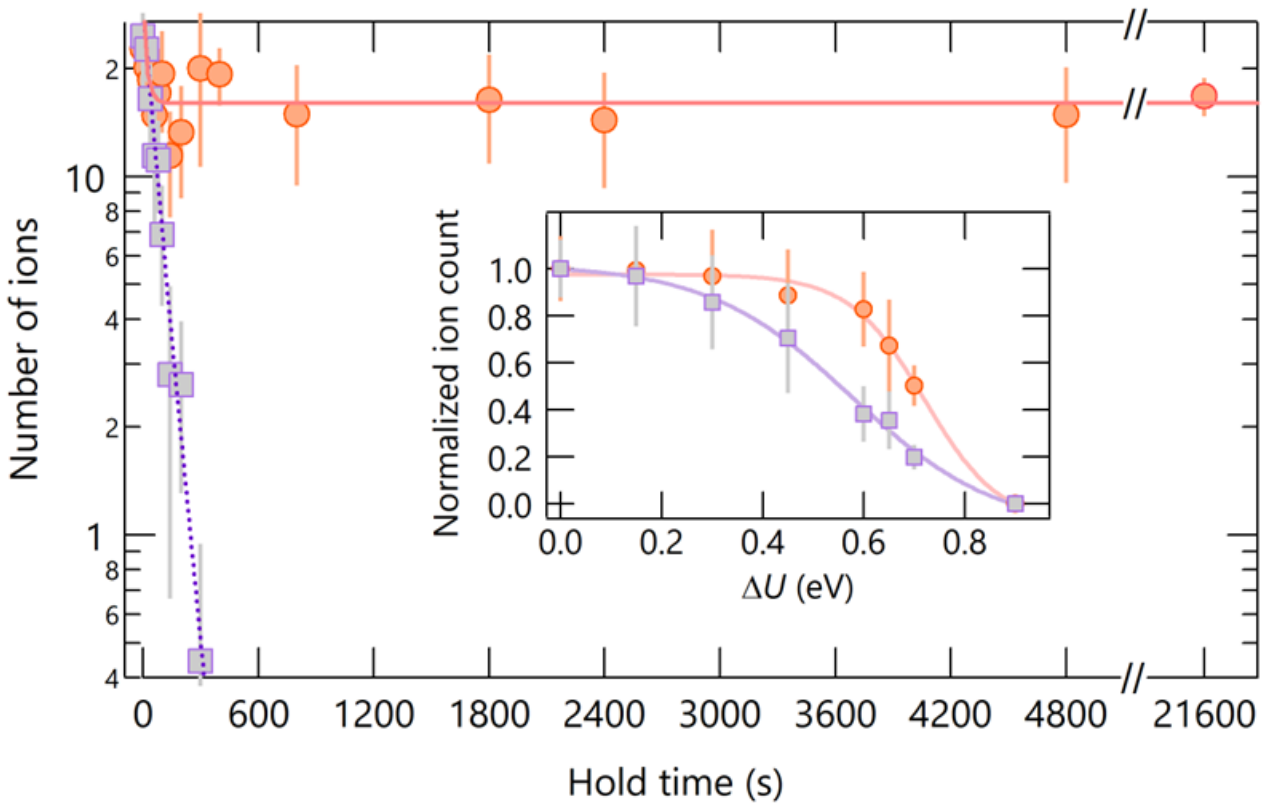

FIG. 1. The number of ions detected after holding the ions in the ion trap for different values of hold times, either in the absence (squares) or presence (circles) of the ultracold atoms. The dotted and solid lines are fits to the model developed for the respective cases. The number of ions reach a steady state value in the presence of ultracold atoms. Inset: Normalized ion count when the trap depth is reduced by $\Delta U$ just prior to extraction of the ions for detection, with ions held either in absence (squares) or presence (circles) of ultracold atoms. The higher ion count in the presence of ultracold atom is a signature of ion cooling by atoms. The sigmoidal curves in the Inset are guides to the eye.

The trajectory of a single ion trapped in a rf Paul trap is separable into two parts [26]: (a) the micromotion, which is the fast motion in sync with the rf drive and whose amplitude is proportional to the rf voltage at the location of the ion. In an LPT, the rf amplitude, and hence the micromotion, is zero on the axis and increases radially outwards. (b) The secular motion (macromotion), which is the slower motion in the effective dynamical potential created by the rf voltage. The secular motion determines the size of the ion's orbit and is a measure of the kinetic energy (i.e. motional temperature) of the ion – smaller the orbit, smaller is the ion's kinetic energy. Cooling an ion is therefore tantamount to reducing the size of the secular orbit [10,26]. Under realistic experimental conditions, the trajectory of an ion expands, i.e. the kinetic energy of the ions increases, due to residual background gas induced collisional heating (at rate $h_b$) [10,11]. In addition, if there are multiple ions in the trap such that the Coulomb interaction is non-negligible, the trajectories expand due ion-ion rf heating (at rate $h_{ii}$) [16,27]. If unchecked, the ion trajectories expand indefinitely eventually leading to their loss from the LPT – this sets the lifetime of the ions in absence the ultracold atoms. The presence of ultracold Cs atoms at the center of the LPT leads to sympathetic cooling (at rate $c_{ia}$) due to elastic [10,12] as well as RCE collisions [11] and arrests the expansion of the ion trajectories resulting in increased lifetime. While elastic collisions reduce only a small fraction of the ion's energy per ion-atom collisions, RCE collisions in homonuclear ion-atom collisions can remove the entire kinetic energy in a single collision, thus providing a very efficient cooling mechanism for the $Cs^+$-Cs case. The competition between the heating and cooling rates determines the lifetime of the ions in the LTP and a steady state is realized when the cooling rate $c_{ia}$ overcomes the total heating rate $h_{tot} = h_b + h_{ii}$.

The background gas pressure, and thus $h_b$, is kept fixed for all experiments reported here. We tune $h_{ii}$ by changing the number of ions initially loaded ($N_i$) in the LPT. In general, $h_{ii}$ has a complicated non-monotonic dependence [16] on the thermal energy ($k_B T$) and ion-ion Coulomb interaction energy $e^2/4\pi\epsilon_0 a$, where $a$ is the Wigner-Seitz radius and is related to the ion density $n$ by $4\pi n a^3/3 = 1$. However, our experiments, with $N_i < 300$ in an LPT of large volume $V$, operate in a restricted regime of temperature (tens to few hundred Kelvin) and low ion density, where $h_{ii}$ monotonically increases with $n$ [16], and hence $N_i$, and is smaller than $h_b$ i.e. $h_{ii} \ll h_b$. Under these conditions, in the absence of ultracold atoms, we model the evolution of $N(t)$ through a rate equation:

$$\frac{dN}{dt} = -\gamma_b N - \beta_{ii} \int n^2 d^3r \qquad (1)$$

where $\gamma_b$ and $\beta_{ii}$ are the loss rate coefficients due to $h_b$ and $h_{ii}$, respectively. Assuming that $n$ remains constant, the second term is approximated as $\beta_{ii} \int n^2 d^3r \approx \beta_{ii} N^2/V$ and Eq. (1) takes the form of a Bernoulli differential equation, the solution for which is:

$$N(t) = \frac{N_i e^{-\gamma_b t}}{1 + \frac{\beta_{ii}/V}{\gamma_b} N_i (1 - e^{-\gamma_b t})} \qquad (2)$$

In Fig. 1, we show the fit (dotted line) of the data in absence of ultracold atoms to Eq. (2) with fitting parameters

$\gamma_b \approx 0.013(4)\ \mathrm{s}^{-1}$ and $\beta_{ii}/V \approx 1.2(3) \times 10^{-4}\ \mathrm{s}^{-1}$. From these, we estimate $h_b \approx \gamma_b U \sim 0.01$ eV/s and $h_{ii}/N \approx (\beta_{ii}/V)U \sim 10^{-4}$ eV/s, where $U \approx 0.9$ eV is the ion trap depth determined from ion trajectory simulations.

To model the ion cooling in the presence of ultracold atoms, we add a positive term $+\beta_{ia}\int n\rho_a d^3r \approx +\beta_{ia} n N_a$ on the right-hand side of Eq. (1). Here, $\rho_a$ and $\beta_{ia}$ represent the atomic density and ion-atom collision rate coefficient, respectively. Since the atomic cloud is much smaller than the ion cloud, we assume the ion density $n \approx N/V$ to be constant over the region of the atomic cloud, such that the term reduces to $+\beta_{ia}(N/V)N_a \equiv \gamma_{ia}N$, where $N_a$ is the number of atoms which remains constant. With this additional term to represent the cooling due to ultracold atoms, the solution is easily obtained by replacing $-\gamma_b \to (-\gamma_b + \gamma_{ia})$ in Eq. (2). Defining $N_s = (\gamma_{ia} - \gamma_b)/(\beta_{ii}/V)$, the solution can be written in a compact form as:

$$N(t) = \frac{N_s}{1 + \left(\frac{N_s}{N_i} - 1\right) e^{(\gamma_b - \gamma_{ia})t}} \qquad (3)$$

In Fig. 1, we show the fit (solid line) of the data (circles) in the presence of ultracold atoms to Eq. (3) with fitting parameter $\gamma_b - \gamma_{ia} \approx -0.04(2)\ \mathrm{s}^{-1}$. Using this along with $\gamma_b \approx 0.013(4)\ \mathrm{s}^{-1}$, we estimate the cooling rate $c_{ia} \approx \gamma_{ia} U \sim 0.05$ eV/s. Since $|c_{ia}| > |h_{tot}|$, ions are effectively cooled and $N(t)$ reaches a steady state $N_s$. This non-zero steady state $N_s$ is consistent with Eq. (3) in the limit $t \to \infty$ when $\gamma_{ia} > \gamma_b$. However, when $\gamma_{ia} < \gamma_b$, $N(t) \to 0$ as $t \to \infty$; consistent with the experimental observation without ultracold atoms.

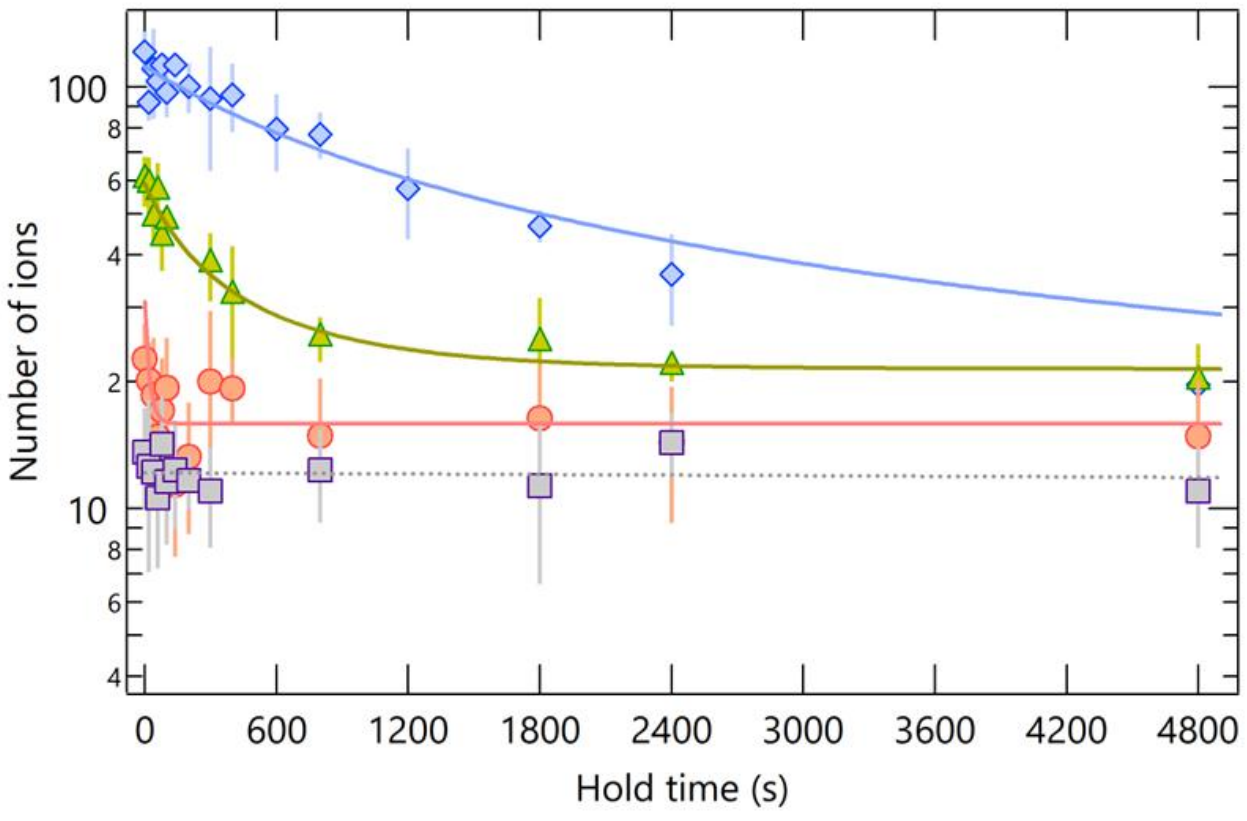

FIG. 2. The number of ions (symbols) plotted as a function of the hold time for different values of initial ion count, while keeping the number of ultracold atoms the same in all cases. The solid lines are fits to Eq. (3). For initial ion count $\gtrsim 22$, the number of ions converges to a steady state value of $N_s \approx 19 \pm 3$ after an initial phase of decay, while no decay is observed for initial ion count $\lesssim$ 22 as depicted by the square symbols.

To test robustness of the steady state, we perform experiments at different values of $N_i$ which correspond to different values of $h_{ii}$. In Fig. 2, we plot $N(t)$ vs. $t$ in the presence of ultracold atoms for different values of initial loadings $N_i$. The initial decay in $N(t)$ suggests a phase where $h_{ii}$ competes with $c_{ia}$ but as later times, when $N(t)$ has decreased sufficiently, the $c_{ia}$ dominates over $h_{ii}$ leading to a steady state. We observe that the value of $N(t)$ converges $N_s \approx 19 \pm 3$ as $t \to \infty$ for all $N_i > N_s$. For $N_i < N_s$, no decay in $N(t)$ is observed (squares in Fig. 2) implying that the cooling rate $c_{ia}$ is sufficient to overcome the heating rate $h_{tot}$ and thus completely arrest the expansion of the ion trajectories. Remarkably, several tens of ions can be held in the LPT in steady state.

In our model, $c_{ia} \approx \gamma_{ia} U$ where $\gamma_{ia} = (\beta_{ia}/V)N_a$, resulting in $c_{ia} \propto N_a$. We expect $c_{ia}$ to increase with $N_a$ because of the enhanced cooling capacity of the atomic cloud. To validate this, we perform experiments at different values of $N_a$. We load the LPT with ~25 ions and hold them with ultracold atoms for a fixed hold time of 30 minutes. Figure 3(a) shows the fraction of remaining ions for different values of $N_a$ ranging from $\sim 10^6$ to $\sim 10^8$ while the atomic cloud's diameter is kept almost constant (deviations <20%). We observe that the fraction of remaining ion increases linearly with $N_a$, which is a result of increase in values of $c_{ia}$ due to the increase in ion-atom collision rate. Further, we expect the ion cooling rate to be the maximum when the ultracold atomic cloud is placed precisely at the center of the LPT where the ion micromotion is the minimum. We test this experimentally by displacing the atomic cloud radially in the LPT and recording the fraction of ions remaining in the LPT after a hold time of 30 minutes. As shown in Fig. 3(b), the fraction of ions remaining is the highest when the atomic cloud is placed at the center of the LPT implying the ion cooling is most effective at the center.

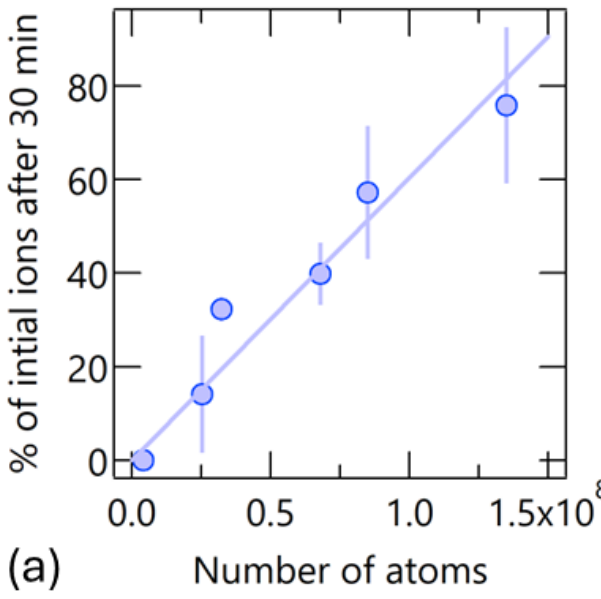

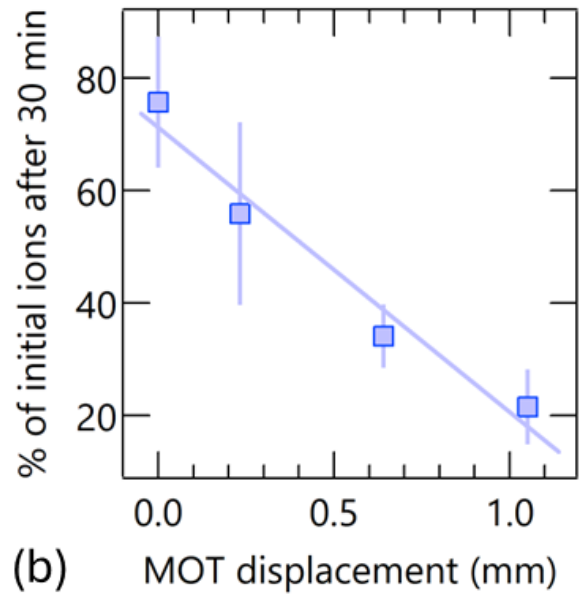

FIG. 3. (a) The fraction of initial ions that remain in the ion trap after 30 minutes when held in the presence of ultracold atomic clouds containing different number of atoms. The fraction of remaining ions increases with the number of atoms due to enhanced ion-atom collision-induced cooling. (b) The fraction of remaining ions decreases as the ultracold atomic cloud in the MOT is displaced radially from the center of the ion trap.

We perform additional experiments to rule out possible alternative explanations for the steady state ion count. (*i*) To rule out the possibility that $Cs^+$ ions might be formed spontaneously in a MOT, we keep the MOT and LPT on for 80 minutes without loading with any ion i.e. the 505-nm photo-ionization light is not applied. We do not detect any ions after 80 minutes, confirming that $Cs^+$ is not formed spontaneously. (*ii*) To rule out the possibility of a fortuitous LPT operating parameters leading to extended ion lifetimes, we perform experiments at different combinations of $(V_{\text{rf}}, \Omega_{\text{rf}})$ viz. $(65\text{ V}, 332\text{ kHz})$, $(100\text{ V}, 412\text{ kHz})$, $(130\text{ V}, 470\text{ kHz})$ and $(150\text{ V}, 505\text{ kHz})$, while keeping the Mathieu stability parameter $q\ (\propto V_{\text{rf}}/\Omega_{\text{rf}}^2)$ fixed. We observe the emergence of long-lived steady state in all cases when ions are held in the presence of $\sim 10^8$ atoms. The value of $N_s$ increases with the trap depth ($\propto qV_{\text{rf}}$) and is the highest, i.e. $N_s \sim 60$, for the $(150\text{ V}, 505\text{ kHz})$ case.

The long trap lifetime of atoms and ions in a hybrid trap permits accumulation of collisions statistics over $\sim 10^4$ ion-atom encounters (at our atomic density) and is therefore important for the studies of rare chemical reactions and inelastic processes. As an example, we consider the rate $(R_3)$ of formation of stable $\text{Cs}_2^+$ molecular ions by the ion-atom-atom three-body recombination [28,29] process viz. $\text{Cs}^+ + \text{Cs} + \text{Cs} \rightarrow \text{Cs}_2^+ + \text{Cs}$. Stable $\text{Cs}_2^+$, if formed, would either be lost from the LPT or be detected via time-of-flight (ToF) mass spectrometry. We do not detect any ions at the expected ToF corresponding to the mass of $\text{Cs}_2^+$ and, as shown above, there is no loss of ions at long hold times. The 6 hours ($\equiv \tau$) of stable $\text{Cs}^+$ trapping with Cs atom thus provides an upper limit for $R_3$ i.e. $R_3 \ll (1/6)\text{hr}^{-1}$. Assuming that at least one ion always remains within the atomic cloud, we have $R_3 \approx k_3 n_a^2$, where $n_a (= 1.8 \times 10^{10}\text{cm}^{-3})$ is the atomic density and $k_3$ is the $\text{Cs}_2^+$ formation rate constant. We thus put an upper bound on $k_3$: $k_3 \ll 1.4 \times 10^{-25}\text{cm}^6\text{s}^{-1}$ for collision energies $\mathcal{O}(10^{-2})$ eV. In the future, it may be possible to determine whether larger molecular ions such as trimers, tetramers etc. are be formed in ion-atom hybrid systems and whether they are stable.

In summary, we experimentally demonstrate a long-lived steady state of ensembles of trapped ions with ensembles of localized ultracold atoms placed precisely at the center of the ion trap. We observe that background heating and ion-ion rf heating competes with the ion-atom collision-induced cooling, and a steady state emerges when the cooling rate overcomes the heating rate. The atomic cloud may be regarded as colder system with a finite heat capacity, where the temperature of the ions is expected to remain higher than the atoms because of the finite spatial extent of the atomic cloud [10,27]. Future experiments could be undertaken to determine whether the ions organize into a Coulomb crystal, especially at higher atomic densities possible with a spatially dark MOT [30]. The emergence of threshold behavior and a stable steady state underscore the system's potential as a versatile platform for exploring non-equilibrium statistical mechanics and relaxation dynamics. Notably, the persistence of this steady state opens avenues for probing processes characterized by low cross-sections. Furthermore, prolonged collisions with atoms lead to cooling not only the center-of-mass motion but also the internal degrees of freedom of molecular ions – this could potentially advance the study of quantum-state-resolved cold chemistry.

**Acknowledgements:** We acknowledge support from the Department of Atomic Energy, Government of India under Project Identification No. RTI4012 and the National Quantum Mission, Government of India.

## Supplementary Material

### A. Ultracold Cs atoms in a magneto-optical trap:

Our vacuum apparatus [23] consists of two sections, the 2D MOT chamber and the 3D MOT chamber, connected by a differential pumping tube. The 2D MOT chamber contains a cesium dispenser source which is operated at a relatively low current of ~2.8 A to maintain a pressure below $\leq 1 \times 10^{-8}$ mbar in the 2D MOT chamber. The pressure in the 3D MOT chamber is maintained at $\leq 1 \times 10^{-11}$ mbar. The laser beams for the 2D MOT and the 3D MOT have the same frequencies: the cooling laser beam is 10 MHz red detuned from the $6s_{1/2}(F = 4) \rightarrow 6p_{3/2}(F' = 5)$ transition, while the repumping laser beam is on resonance with the $6s_{1/2}(F = 3) \rightarrow 6p_{3/2}(F' = 4)$ transition. The cooling (repumping) laser intensities per beam for the 2D MOT and 3D MOT are 80 and 11.6 mW/cm$^2$ (6 and 4 mW/cm$^2$), respectively. The 2D MOT is formed by two pairs of orthogonal, retro-reflected laser beams which intersect at the center of the 2D MOT chamber where a magnetic field gradient of ~7.7 Gauss/cm is applied using electromagnets. The typical diameter of the 2D MOT laser beams is ~25 mm but apertures placed in the laser beam path are used to change the beam diameter and hence the MOT loading rate when needed – this allows control over the number of atoms loaded in the 3D MOT. The 3D MOT is formed by three pairs of mutually orthogonal, retro-reflected laser beams which intersect at the center of the 3D MOT chamber which also happens to be the center of the linear Paul trap. A magnetic field gradient of ~14 Gauss/cm is used for the 3D MOT. Three pairs of electromagnetic coils are placed around the 3D MOT chamber to cancel the ambient magnetic field. The same coils are used to apply calibrated magnetic field to shift the position of the 3D MOT in control experiments. The atoms in the 3D MOT are detected using fluorescence imaging – both on a photodiode (for atom number determination) and a CCD camera (for cloud size determination). We typically load around $1.8 \times 10^8$ atoms in the 3D MOT with a density of around $2 \times 10^{10}$ cm$^{-3}$. Mechanical optical shutters placed in the path of the laser beams are used to control the atom loading sequence in the experiment.

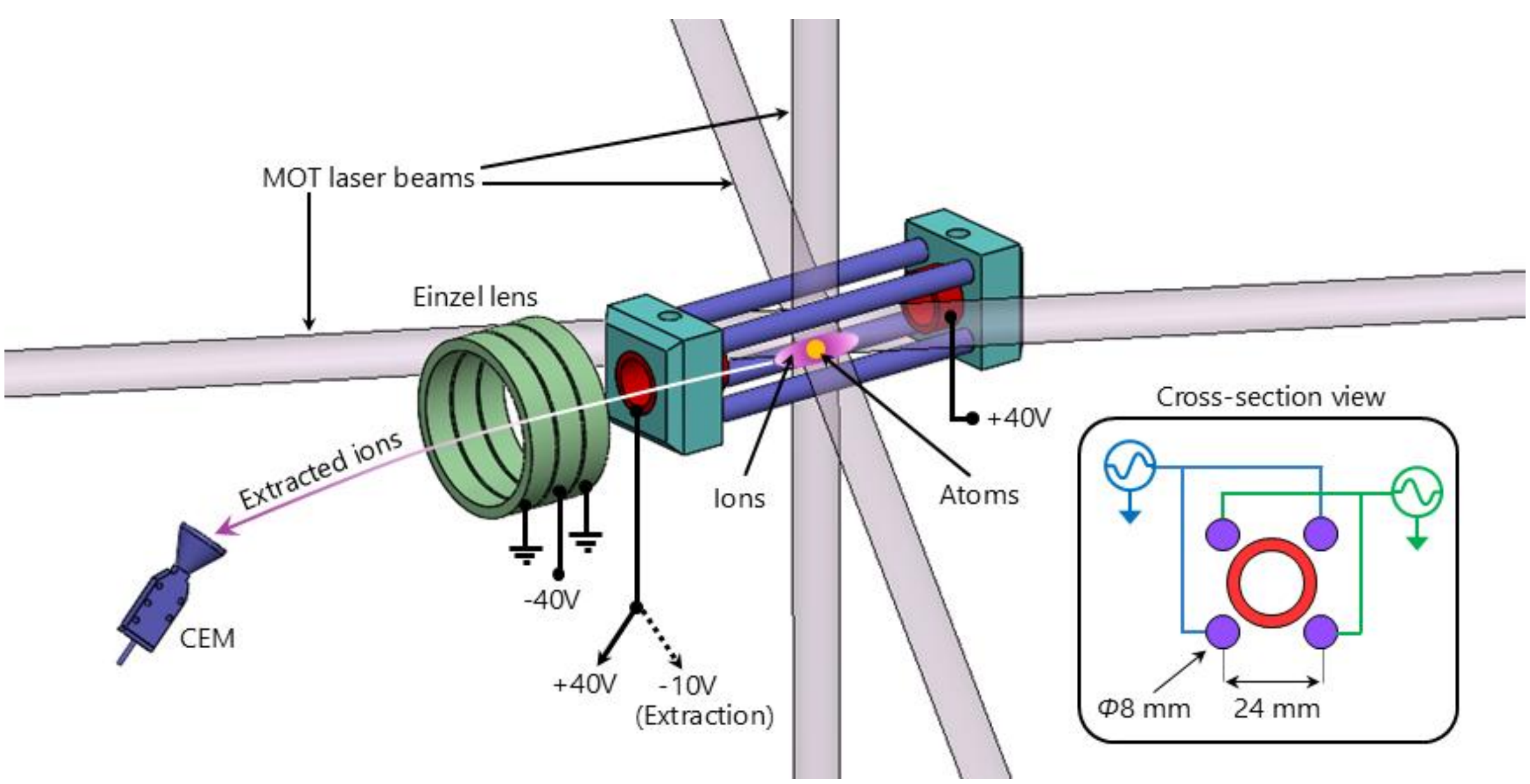

Fig. 1S. Schematic diagram showing the LPT, Einzel lens, CEM and MOT laser beams. Inset: cross-section view of the LPT and the rf voltage configuration.

**B. Cs⁺ ions in a linear Paul trap:**

The center of the LPT is located at the center of the 3D MOT chamber by design. As shown in Fig. 1S, the LPT consists of four rods (diameter 8 mm each) and two annular endcaps. The four rods are placed in a square geometry such that their center-to-center distance is 24 mm, leaving a gap of 16 mm which allows the 3D MOT laser beams to pass through. Radiofrequency (rf) voltages are applied to the rod electrodes such that diagonal electrodes have the same voltage, while adjacent electrodes have voltages 180° out of phase. The rf voltage is obtained by amplifying a sinusoidal signal sourced from a digital signal generator which can be dynamically controlled. We typically apply rf voltages with amplitude of 130 V and frequency of 470 kHz to the rod electrodes. The endcaps are spaced by 88 mm and are annular in shape with inner and outer diameters of 16 mm and 22 mm, respectively. The 16 mm clearance in the endcap allows ions to be extracted axially and detected on a channel electron multiplier (CEM) located at a distance ~262 mm from the center of the LPT. We apply dc voltage of +40 V to both the endcaps for trapping ions. To detect the ions, we switch one of the endcap voltages to -10 V within a microsecond (see Fig. 1S). The extracted ions pass through an Einzel lens set at -40 V dc before being detected in the CEM. The arrival time of the ions in the CEM depends on the mass of the ions as in a typical time-of-flight (TOF) mass spectrometer (MS). The TOFMS has sufficient mass resolution to distinguish between $\mathrm{Cs}^+$ and $\mathrm{Cs}_2^+$ ions.

To characterize the LPT, we measure the radial and axial secular frequencies of the ions using the parametric resonant quenching method. To measure the radial secular frequency, we apply an additional rf drive of variable frequency ($f_{tickle}$) with low amplitude (~250 mV) to the four rod electrodes and measure the number of trapped ions as $f_{tickle}$ is tuned. The number of ions decreases when $f_{tickle}$ matches twice the radial secular frequency (see Fig. 2S). To measure the axial secular frequency, we apply the additional rf voltage to one of the endcaps and repeat the procedure. The measured radial and axial secular frequencies are $\omega_r = 2\pi \times 37$ kHz and $\omega_a = 2\pi \times 12$ kHz, respectively. We additionally carry out simulations using SIMION and find the excellent agreement with the measured secular frequencies. To estimate the trap depth of the LPT we rely on SIMION simulations as described in Ref. [25]. We displace the ions radially from centre of the LPT and determine the radial distance ($r_{max}$) beyond which ions start escaping from the trap. The trap depth $U$~0.9 eV is then estimated from the expression $U \approx m\omega_r^2 r_{max}^2/2$.

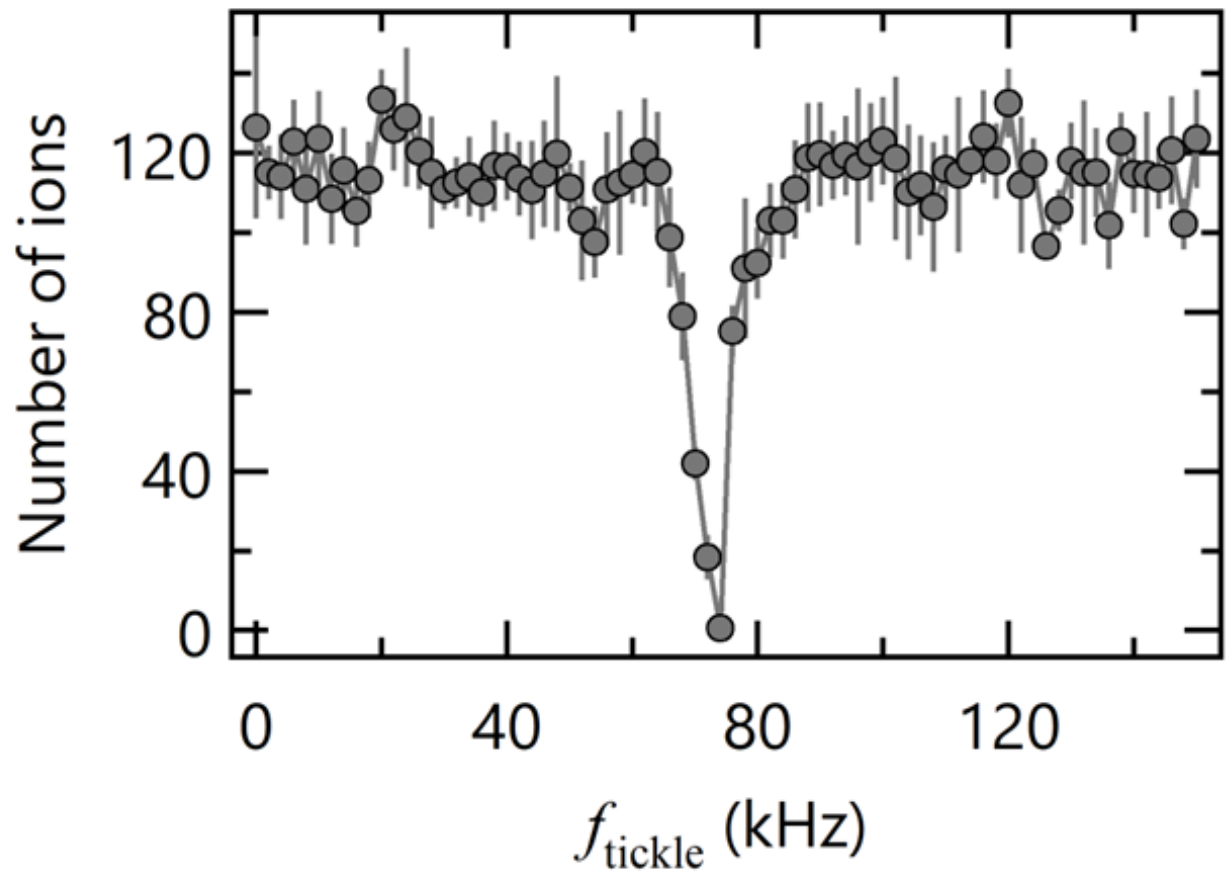

Fig. 2S. Number of ions plotted against the additional radial drive frequency $f_{tickle}$.

### C. Temperature of $Cs^+$ ions:

We note that $Cs^+$ ions cannot be laser-cooled and do not have transitions that can be readily addressed with lasers. Therefore, the typical approaches of ion temperature determination based on fluorescence or sideband thermometry are not applicable to $Cs^+$ ions. In this situation, one way to estimate [25] the maximum ion temperature $T_{max}$ is to assume that the ion speeds ($v$) follow a Maxwell-Boltzmann (MB) distribution $f(v)$ and have an upper cut-off $v_{max} \sim 1140$ m/s determined by the trap depth $U \sim 0.9$ eV. Here, we used the expression $U = mv_{max}^2/2$ to estimate $v_{max}$. Thus, the MB distribution should be such that all (say, >99.99%) ions have speeds $< v_{max}$. The MB distribution $f(v)$ that satisfies this condition has a temperature $T_{max} \sim 900$ K and most probable speed $v_{mp} \sim 335$ m/s, which corresponds to a most probable ion energy $v_{mp} \sim 78$ meV. The temperature of ions which remain trapped for extended periods, say >1 s, are expected to be significantly lower compared to $T_{max}$. Our earlier experiments have estimated the ion temperature to be ~350 K under similar conditions [11]. The temperature decreases after the ions are cooled by the ultracold atoms.

In our experiments, the $Cs^+$ ions are created by photoionization of ultracold Cs atoms in the $6p_{3/2}$ state using light at 505-nm light. The choice of 505-nm causes near-threshold ionization of Cs to produce $Cs^+$ ions with nearly zero kinetic energy (the light wavelength required for threshold ionization is 508.2 nm). However, the $Cs^+$ ions are produced at different locations in the ion trap with a distribution that mimics the size of the ultracold atomic cloud which has a radius of $\sim 0.9$ mm. Thus, the ions have an initial potential energy which gets converted to kinetic energy as the ions move towards the center of the trap. This leads to a non-zero initial ion temperature. The ion trap depth at a radial distance $r \sim 0.9$ mm from the center of the LPT is ~30 meV, where we have used the expression $U(r) = m\omega_r^2 r^2/2$. The initial energy of ~30 meV amounts to an initial ion temperature of ~350 K. As the ions collide with the ultracold atoms at the center of the LPT, they lose part of the kinetic energy and become colder which manifests in the increase in the lifetime of the ions in the LPT.

### D. Experimental sequence:

The experimental sequence to measure the lifetime of the trapped ions is the following. We first unblock the light for the 3D MOT and allow the 3D MOT to load till the atom number saturates. The ion trap is then turned on by switching on the rf and dc voltages. Ions are then created by a short pulse of 505-nm light, a fraction of which are trapped in the LPT. After this, the 3D MOT light is kept on or switched off depending on whether we want to hold the ions with or without the ultracold atoms. The ions are held in the LPT for a predetermined hold-time, after which the dc voltage on the endcap closer to the CEM is switched to -10 V to extract the ions and detect them on the CEM. The rf voltages are kept on until the ions hit the CEM. Individual ions hitting the CEM give individual current pulses of width ~10 ns which are recorded as individual voltage pulses on a digital storage oscilloscope and counted in post-processing by settings an appropriate discriminator voltage. The number of trapped ions is kept low (below 300) to avoid overlapping pulses and to avoid detector saturation effects. At each hold-time, the above sequence is repeated several times. The average ion count and the standard deviation are determined and plotted in the main text. To avoid biases arising small drifts in experimental parameters during the several hours to days of data collection, the with and without MOT data are taken in an interspersed manner.

---

[1] M. Tomza, K. Jachymski, R. Gerritsma, A. Negretti, T. Calarco, Z. Idziaszek, and P. S. Julienne, Cold hybrid ion-atom systems, Rev. Mod. Phys. **91**, 035001 (2019).

[2] A. T. Grier, M. Cetina, F. Oručević, and V. Vuletić, Observation of cold collisions between trapped ions and trapped atoms, Phys. Rev. Lett. **102**, 223201 (2009).

[3] C. Zipkes, S. Palzer, L. Ratschbacher, C. Sias, and M. Köhl, Cold heteronuclear atom-ion collisions, Phys. Rev. Lett. **105**, 133201 (2010).

[4] S. Schmid, A. Härter, and J. Hecker Denschlag, Dynamics of a cold trapped ion in a Bose-Einstein condensate, Phys. Rev. Lett. **105**, 133202 (2010).

[5] F. H. J. Hall, M. Aymar, N. Bouloufa-Maafa, O. Dulieu, and S. Willitsch, Light-assisted ion-neutral reactive processes in the cold regime: Radiative molecule formation versus charge exchange, Phys. Rev. Lett. **107**, 243202 (2011).

[6] P. Weckesser, F. Thielemann, D. Wiater, A. Wojciechowska, L. Karpa, K. Jachymski, M. Tomza, T. Walker, and T. Schaetz, Observation of Feshbach resonances between a single ion and ultracold atoms, Nature **600**, 429 (2021).

[7] S. Dutta and S. A. Rangwala, Nondestructive detection of ions using atom-cavity collective strong coupling, Phys. Rev. A **94**, 053841 (2016).

[8] K. Ravi, S. Lee, A. Sharma, G. Werth, and S. A. Rangwala, Cooling and stabilization by collisions in a mixed ion-atom system, Nat. Commun. **3**, 1126 (2012).

[9] I. Sivarajah, D. S. Goodman, J. E. Wells, F. A. Narducci, and W. W. Smith, Evidence of sympathetic cooling of $Na^+$ ions by a Na magneto-optical trap in a hybrid trap, Phys. Rev. A **86**, 063419 (2012).

[10] S. Dutta, R. Sawant, and S. A. Rangwala, Collisional Cooling of Light Ions by Cotrapped Heavy Atoms, Phys. Rev. Lett. **118**, 113401 (2017).

[11] S. Dutta and S. A. Rangwala, Cooling of trapped ions by resonant charge exchange, Phys. Rev. A **97**, 041401(R) (2018).

[12] S. Haze, M. Sasakawa, R. Saito, R. Nakai, and T. Mukaiyama, Cooling Dynamics of a Single Trapped Ion via Elastic Collisions with Small-Mass Atoms, Phys. Rev. Lett. **120**, 043401 (2018).

[13] T. Feldker, H. Fürst, H. Hirzler, N. V. Ewald, M. Mazzanti, D. Wiater, M. Tomza, and R. Gerritsma, Buffer gas cooling of a trapped ion to the quantum regime, Nat. Phys. **16**, 413 (2020).

[14] B. Höltkemeier, P. Weckesser, H. López-Carrera, and M. Weidemüller, Buffer-Gas Cooling of a Single Ion in a Multipole Radio Frequency Trap beyond the Critical Mass Ratio, Phys. Rev. Lett. **116**, 233003 (2016).

[15] Z. Meir, T. Sikorsky, R. Ben-Shlomi, N. Akerman, Y. Dallal, and R. Ozeri, Dynamics of a Ground-State Cooled Ion Colliding with Ultracold Atoms, Phys. Rev. Lett. **117**, 243401 (2016).

[16] S. J. Schowalter, A. J. Dunning, K. Chen, P. Puri, C. Schneider, and E. R. Hudson, Blue-sky bifurcation of ion energies and the limits of neutral-gas sympathetic cooling of trapped ions, Nat. Commun. **7**, 12448 (2016).

[17] I. Rouse and S. Willitsch, Superstatistical Energy Distributions of an Ion in an Ultracold Buffer Gas, Phys. Rev. Lett. **118**, 143401 (2017).

[18] M. Pinkas, O. Katz, J. Wengrowicz, N. Akerman, and R. Ozeri, Trap-assisted formation of atom-ion bound states, Nat. Phys. **19**, 1573 (2023).

[19] H. Hirzler, E. Trimby, R. Gerritsma, A. Safavi-Naini, and J. Pérez-Ríos, Trap-Assisted Complexes in Cold Atom-Ion Collisions, Phys. Rev. Lett. **130**, 143003 (2023).

[20] L. Ratschbacher, C. Sias, L. Carcagni, J. M. Silver, C. Zipkes, and M. Köhl, Decoherence of a single-ion qubit immersed in a spin-polarized atomic bath, Phys. Rev. Lett. **110**, 160402 (2013).

[21] R. Saito, S. Haze, M. Sasakawa, R. Nakai, M. Raoult, H. Da Silva, O. Dulieu, and T. Mukaiyama, Characterization of charge-exchange collisions between ultracold $^{6}$Li atoms and $^{40}$Ca$^{+}$ ions, Phys. Rev. A **95**, 032709 (2017).

[22] A. Mahdian, A. Krükow, and J. Hecker Denschlag, Direct observation of swap cooling in atom-ion collisions, New J. Phys. **23**, 065008 (2021).

[23] B. Rahaman, S. Baidya, and S. Dutta, A versatile apparatus for simultaneous trapping of multiple species of ultracold atoms and ions to enable studies of low energy collisions and cold chemistry, J. Chem. Phys. **160**, 064201 (2024).

[24] Supplementary Material

[25] S. Dutta and S. A. Rangwala, Measurement of collisions between laser-cooled cesium atoms and trapped cesium ions, Phys. Rev. A **102**, 033309 (2020).

[26] F. G. Major, V. N. Gheorghe, and G. Werth, *Charged Particle Traps: Physics and Techniques of Charged Particle Field Confinement* (Springer, 2005).

[27] D. S. Goodman, I. Sivarajah, J. E. Wells, F. A. Narducci, and W. W. Smith, Ion-neutral-atom sympathetic cooling in a hybrid linear rf Paul and magneto-optical trap, Phys. Rev. A **86**, 033408 (2012).

[28] A. Härter, A. Krükow, A. Brunner, W. Schnitzler, S. Schmid, and J. Hecker Denschlag, Single ion as a three-body reaction center in an ultracold atomic gas, Phys. Rev. Lett. **109**, 123201 (2012).

[29] J. Pérez-Ríos and C. H. Greene, Communication: Classical threshold law for ion-neutral-neutral three-body recombination, J. Chem. Phys. **143**, 041105 (2015).

[30] C. G. Townsend, N. H. Edwards, K. P. Zetie, C. J. Cooper, J. Rink, and C. J. Foot, High-density trapping of cesium atoms in a dark magneto-optical trap, Phys. Rev. A **53**, 1702 (1996).